\documentclass{nature}

\usepackage{graphics}
\usepackage{float}
\usepackage{graphicx}
\usepackage{amsfonts}
\usepackage{textcomp}
\usepackage{amssymb}
\usepackage{amsmath}

\title{Dense Packings of the Platonic and Archimedean Solids}

\author{S. Torquato$^{1,2,3,4,5}$ \& Y. Jiao$^5$}

\begin{document}

\maketitle

\begin{affiliations}
 \item Department of Chemistry, Princeton University,
Princeton New Jersey 08544, USA
 \item Princeton Center for Theoretical Science,
Princeton University, Princeton New Jersey 08544, USA
 \item Princeton Institute for the Science and Technology of
Materials, Princeton University, Princeton New Jersey 08544, USA
\item School of Natural Sciences, Institute for
Advanced Study, Princeton New Jersey 08540, USA
 \item Department of Mechanical and Aerospace Engineering,
Princeton University, Princeton New Jersey 08544, USA

\end{affiliations}

\begin{abstract}
Dense packings 
have served as useful models of the structure 
of liquid, glassy and crystal states of matter \cite{Ber65,Za83,Ch00,To02},
granular media \cite{Ed94,To02}, heterogeneous
materials \cite{To02}, and biological systems \cite{Li01,Pu03,Ge08}. Probing the 
symmetries and other mathematical properties of the densest packings is a 
problem of long-standing interest in discrete geometry and number theory \cite{Co98,Co03,Ha05}.
The preponderance of previous work has focused on spherical particles,
and very little is known about dense polyhedral packings. 
We formulate the problem of generating dense packings of 
polyhedra within an adaptive fundamental cell subject to periodic boundary 
conditions as an optimization problem, which we call the Adaptive 
Shrinking Cell (ASC) scheme. This novel optimization problem is solved 
here (using a variety of multi-particle initial configurations) to find 
dense packings of each of the Platonic solids in three-dimensional 
Euclidean space. We find
the densest known packings of tetrahedra, octahedra, dodecahedra
and icosahedra with densities $0.782\ldots$, $0.947\ldots$,
$0.904\ldots$, and $0.836\ldots$, respectively. Unlike the densest
tetrahedral packing, which must be a non-Bravais  lattice
packing, the densest packings of the other non-tiling Platonic
solids that we obtain are their previously known optimal (Bravais)
lattice packings.  Our simulations 
results, rigorous upper bounds that we derive, and theoretical arguments 
lead us to the strong conjecture that the densest packings of the Platonic 
and Archimedean solids with central symmetry are given by their 
corresponding densest lattice packings. This is the analog of Kepler's sphere conjecture for 
these solids. 
\end{abstract}

A large collection of nonoverlapping solid objects (particles) in
$d$-dimensional Euclidean space $\mathbb{R}^d$ is called a
packing. The packing density $\phi$ is defined as the fraction of
space $\mathbb{R}^d$ covered by the particles. A problem that has
been a source of fascination to mathematicians and scientists for
centuries is the determination of the densest arrangement(s) of
particles that do not tile space and the associated maximal
density $\phi_{max}$ \cite{Co98}.
The preponderance of previous work has focused on spherical particles but, 
even for this simple shape, the problem is notoriously difficult. Indeed, 
Kepler's conjecture concerning the densest sphere packing arrangement was 
only proved by Hales in 2005 \cite{Ha05}. 

Attention has very recently turned to
finding the maximal-density packings of nonspherical
particles in $\mathbb{R}^3$, including ellipsoids \cite{Do04}, 
tetrahedra \cite{Co06,Ch08}, 
and  superballs \cite{Ji09}.  Very little is known about the
densest packings of polyhedral particles that do not tile
space, including the majority of the Platonic and Archimedean solids studied by 
the ancient Greeks. The difficulty in obtaining dense packings of 
polyhedra is related to their complex rotational degrees of freedom and to 
the non-smooth nature of their shapes.

\begin{figure}
\begin{center}
\includegraphics[width=15cm,keepaspectratio]{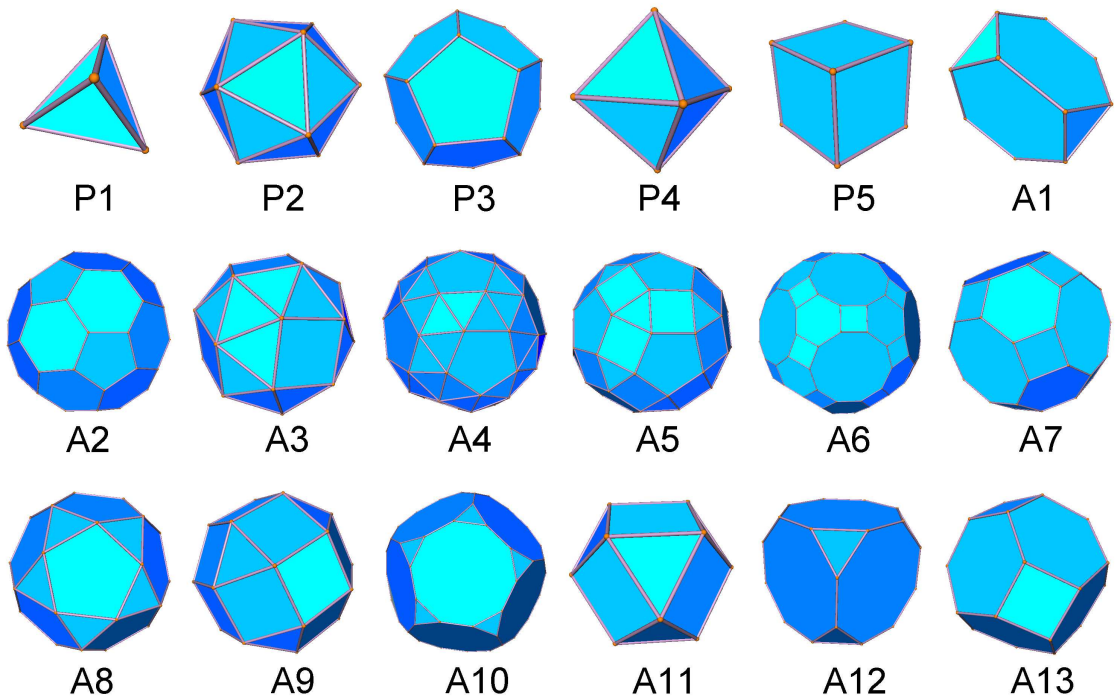} \\
\end{center}
\caption{The five Platonic solids:
tetrahedron (P1), icosahedron (P2), dodecahedron (P3), octahedron (P4) 
and cube (P5). The 13 Archimedean solids: truncated tetrahedron (A1), 
truncated icosahedron (A2),
snub cube (A3), snub dodecahedron (A4), rhombicosidodecahdron (A5), 
truncated icosidodecahdron (A6), 
truncated cuboctahedron (A7), icosidodecahedron (A8),  rhombicuboctahedron (A9),
truncated dodecahedron (A10), cuboctahedron (A11), 
truncated cube (A12), and truncated octahedron (A13). 
Note that the cube (P5) and truncated octahedron (A13) are the only Platonic 
and Archimedean solids, respectively, that tile space.} \label{platonic}
\end{figure}

The Platonic solids  (mentioned in Plato's Timaeus) are convex
polyhedra with faces composed of congruent convex
regular polygons. There are exactly five such solids: the
tetrahedron, icosahedron, dodecahedron, octahedron and cube (see
Fig. \ref{platonic}). An Archimedean solid is a highly symmetric, semi-regular
convex polyhedron composed of two or more types of regular polygons meeting in 
identical vertices. There are thirteen Archimedean solids (see Fig. \ref{platonic}).
Note that the tetrahedron (P1) and truncated tetrahedron (A1) are the only
Platonic and Archimedean solids, respectively, that are not {\it centrally symmetric}. A particle
is centrally symmetric if it has a center $C$ that bisects every
chord through $C$ connecting any two boundary points of the
particle. We will see that this type of symmetry 
plays a fundamental role in determining the nature of the
dense packing arrangements.

Some definitions are in order here. A {\it lattice} $\Lambda$ in $\mathbb{R}^3$ is 
an infinite set of points generated by a set of discrete translation operations 
(defined by integer linear combinations of a basis of $\mathbb{R}^3)$ \cite{Ch00}.
A (Bravais) {\it lattice packing}  is one in which  the centroids of the nonoverlapping
particles are located at the points of $\Lambda$, each oriented in
the same direction. The space $\mathbb{R}^3$
can then be geometrically divided into identical regions $F$ called
{\it fundamental cells}, each of which contains just the centroid
of one particle. Thus, the density of a lattice packing  is given by
\begin{equation}
\phi= \frac{v_{particle}}{\mbox{Vol}(F)},
\end{equation}
where $v_{particle}$ is the volume of a  particle and
$\mbox{Vol}(F)$ is the volume of a fundamental cell.
A {\it periodic} packing of particles is
obtained by placing a fixed nonoverlapping configuration of $N$
particles (where $N\ge 1$) with {\it arbitrary orientations} in
each fundamental cell of a lattice $\Lambda$. Thus, the packing is
still periodic under translations by $\Lambda$, but the $N$
particles can occur anywhere in the chosen cell
subject to the nonoverlap condition. The density of a
periodic packing is given by
\begin{equation}
\phi=\frac{N v_{particle}}{\mbox{Vol}(F)}.
\end{equation}

\begin{figure}
\begin{center}
$\begin{array}{c@{\hspace{0.5cm}}c@{\hspace{0.5cm}}c}\\
\includegraphics[width=5.0cm,keepaspectratio]{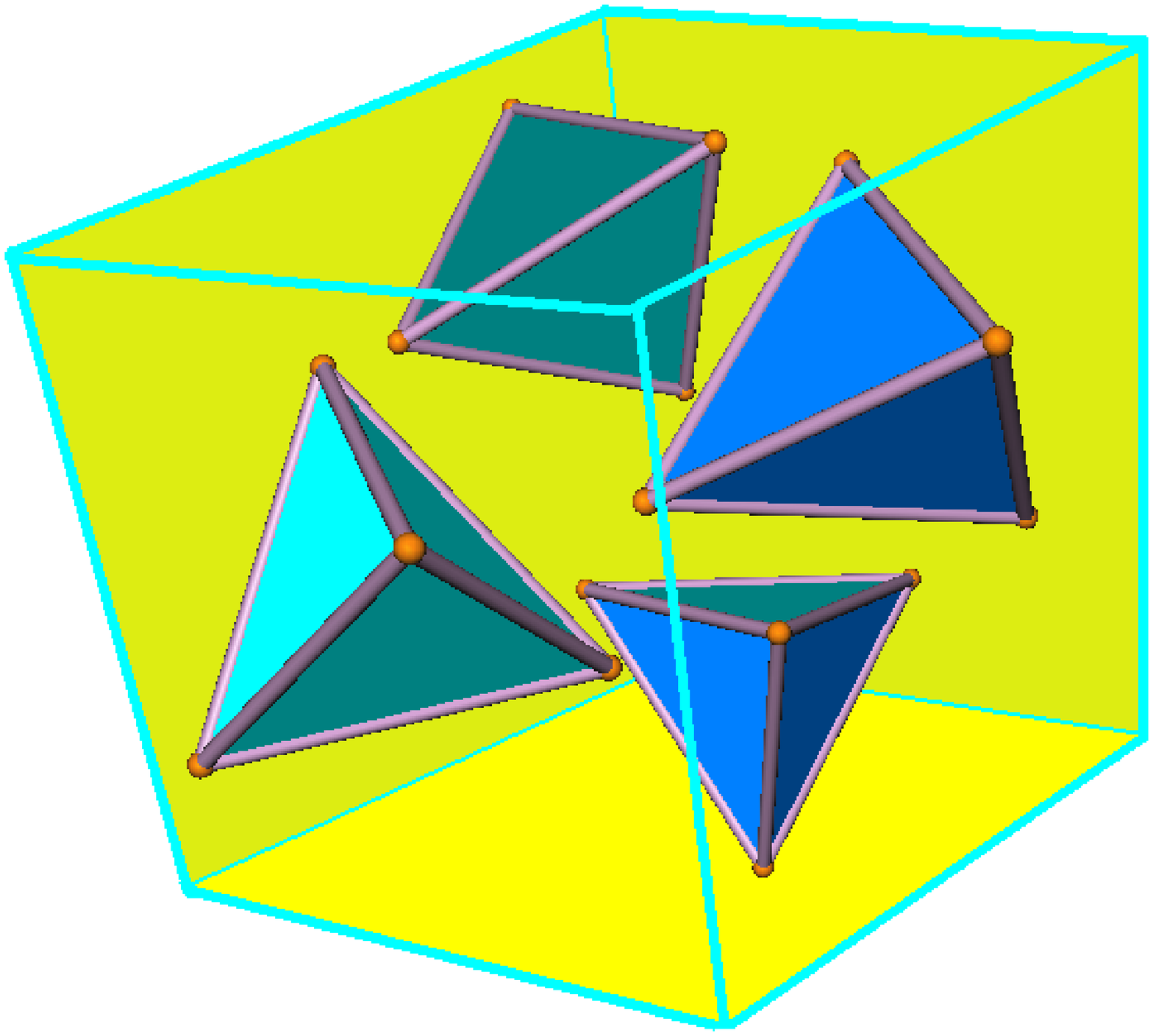} &
\includegraphics[width=5.0cm,keepaspectratio]{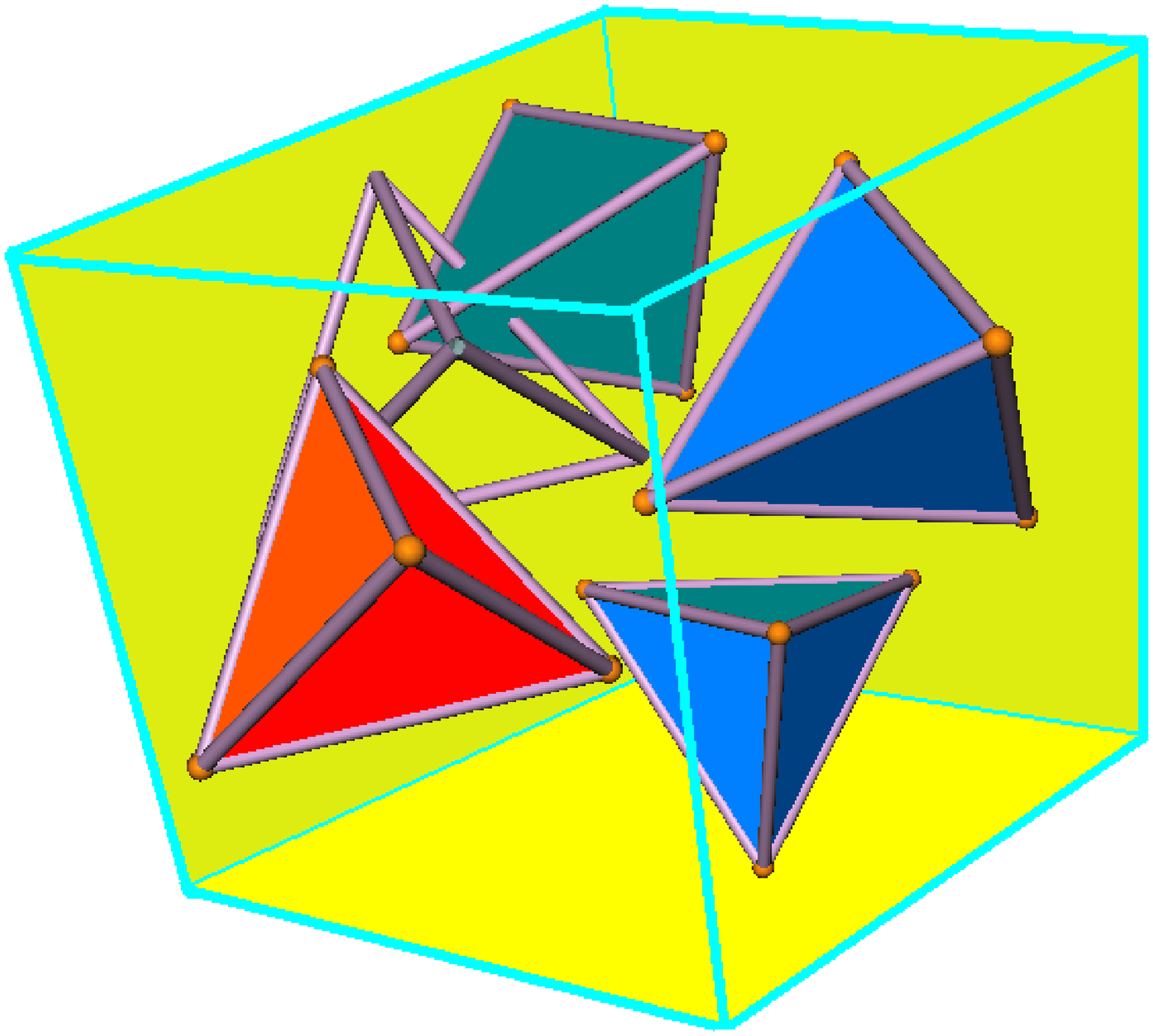} &
\includegraphics[width=5.0cm,keepaspectratio]{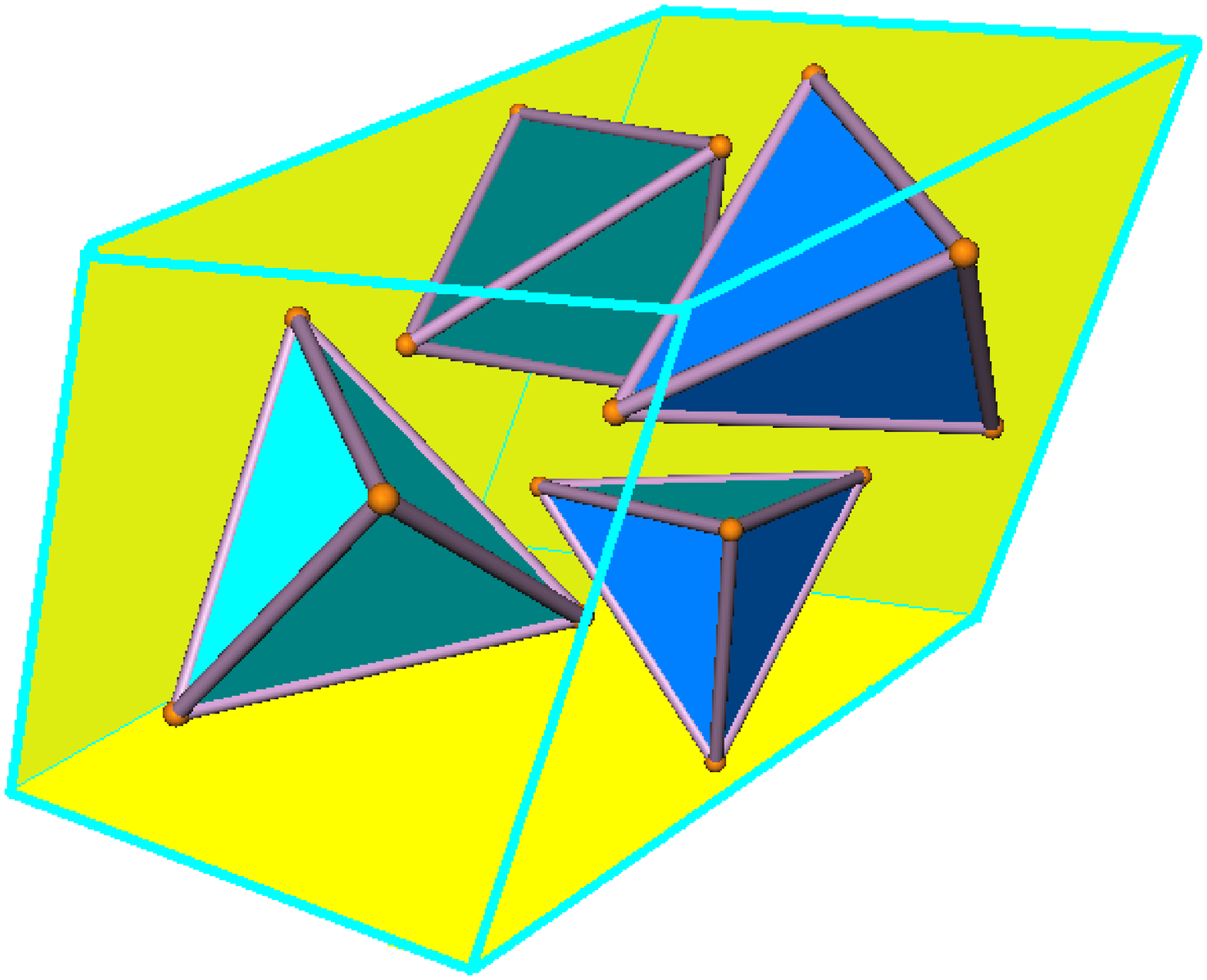} \\
\mbox{\bf (a)} & \mbox{\bf (b)} & \mbox{\bf (c)}
\end{array}$
\end{center}
\caption{ By efficiently exploring the design-variable space (DVS),
which consists of the particle configurational space and the space of lattices 
(due to our use of an adaptive
fundamental cell), the Adapative Shrinking cell (ASC) scheme enables one to find a point in
the DVS in the neighborhood of the starting point that has a
higher packing density than the initial density. The process is
continued until the deepest minimum of the objective function (a
maximum of packing density) is obtained, which could be either a
local or global optimum. Here we show a series of sequential changes of 
a four-particle packing configuration
due to the design variables in the ASC algorithm. (a) An initial configuration of four particles. (b) A trial
move of a randomly selected particle (colored red in this frame)
that is rejected because it overlaps another particle.
This is determined precisely using the separation axis theorem \cite{Ra94}.
(c) A trial move that is accepted, which results
in a deformation and compression (small in magnitude) changing
the fundamental cell shape and size as well as the relative distances between the particles.}
\label{fig_Cell}
\end{figure}

We formulate the problem of generating dense packings of
nonoverlapping polyhedra  within an adaptive fundamental cell 
 subject to periodic boundary conditions as an
optimization problem (see Methods Summary).  We call this optimization scheme the Adaptive
Shrinking Cell (ASC).  
Figure \ref{fig_Cell} illustrates a simple sequence of configuration
changes for a four-particle packing.

Finding the densest packings of regular tetrahedra is
part of the 18th problem in Hilbert's famous set of problems.
 The densest (Bravais) lattice packing of tetrahedra (which
requires all of the tetrahedra to have the same orientations) has
the relatively low density $\phi^{lattice}_{max} =18/49=0.367\ldots$ and
each tetrahedron touches 14 others \cite{Ho70}. Recently, Conway
and Torquato showed that the densest packings of 
tetrahedra must be non-Bravais lattice  packings, and
found packings with density as large as $\phi \approx 0.72$
\cite{Co06}.   
Chaikin, Wang and Jaoshvili experimentally generated jammed
disordered packings of nearly ``tetrahedral" dice with $\phi \approx 0.75$.
Chen \cite{Ch08} has recently discovered a periodic
packing of tetrahedra with $\phi=0.7786\ldots$.
We call this the ``wagon-wheels" packing because the basic
subunits consist of two orthogonally intersecting ``wagon" wheels.
A ``wagon wheel" consists of five contacting tetrahedra packed
around a common edge (see Fig. 1a of Ref. 13).

We begin by solving the ASC scheme to obtain dense packings of tetrahedra
using initial configurations based upon low-density
versions of the aforementioned packings.
Initial conditions based on periodic copies of the wagon-wheels packing with 72
particles per cell
lead to the densest packing of tetrahedra reported to date with $\phi=0.7820021\ldots$
(see Fig. \ref{tetra}). Its lattice vectors and other characteristics are given in
the Supplementary Information. The preference for face-to-face (not vertex-to-face)
contacts and the lack of central symmetry ensure that dense tetrahedral packings
must be non-lattice structures.

\begin{figure}
\begin{center}
$\begin{array}{c@{\hspace{0.1cm}}c@{\hspace{0.1cm}}c@{\hspace{0.1cm}}c}\\
\includegraphics[width=3.5cm,keepaspectratio]{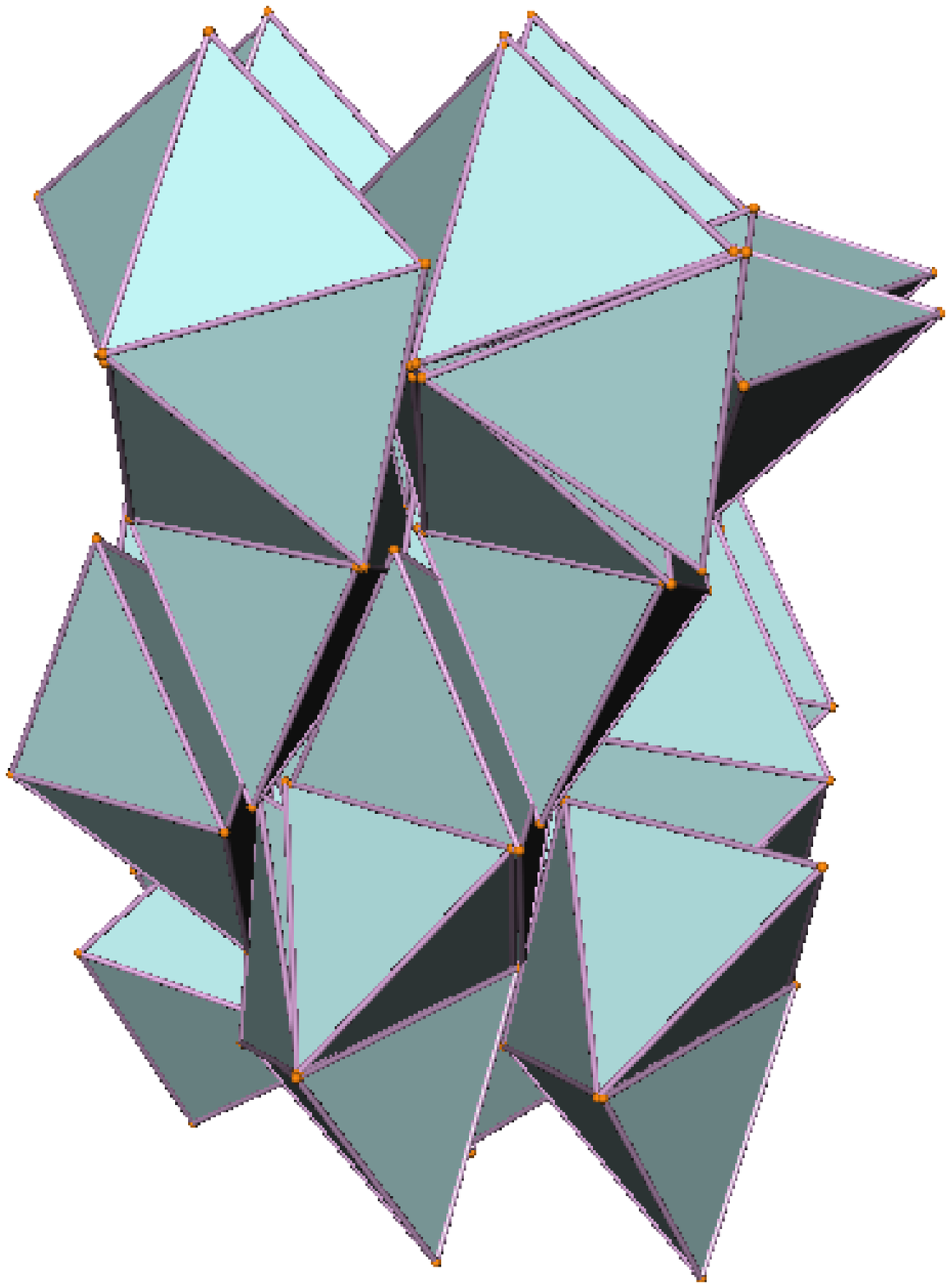} &
\includegraphics[width=4.5cm,keepaspectratio]{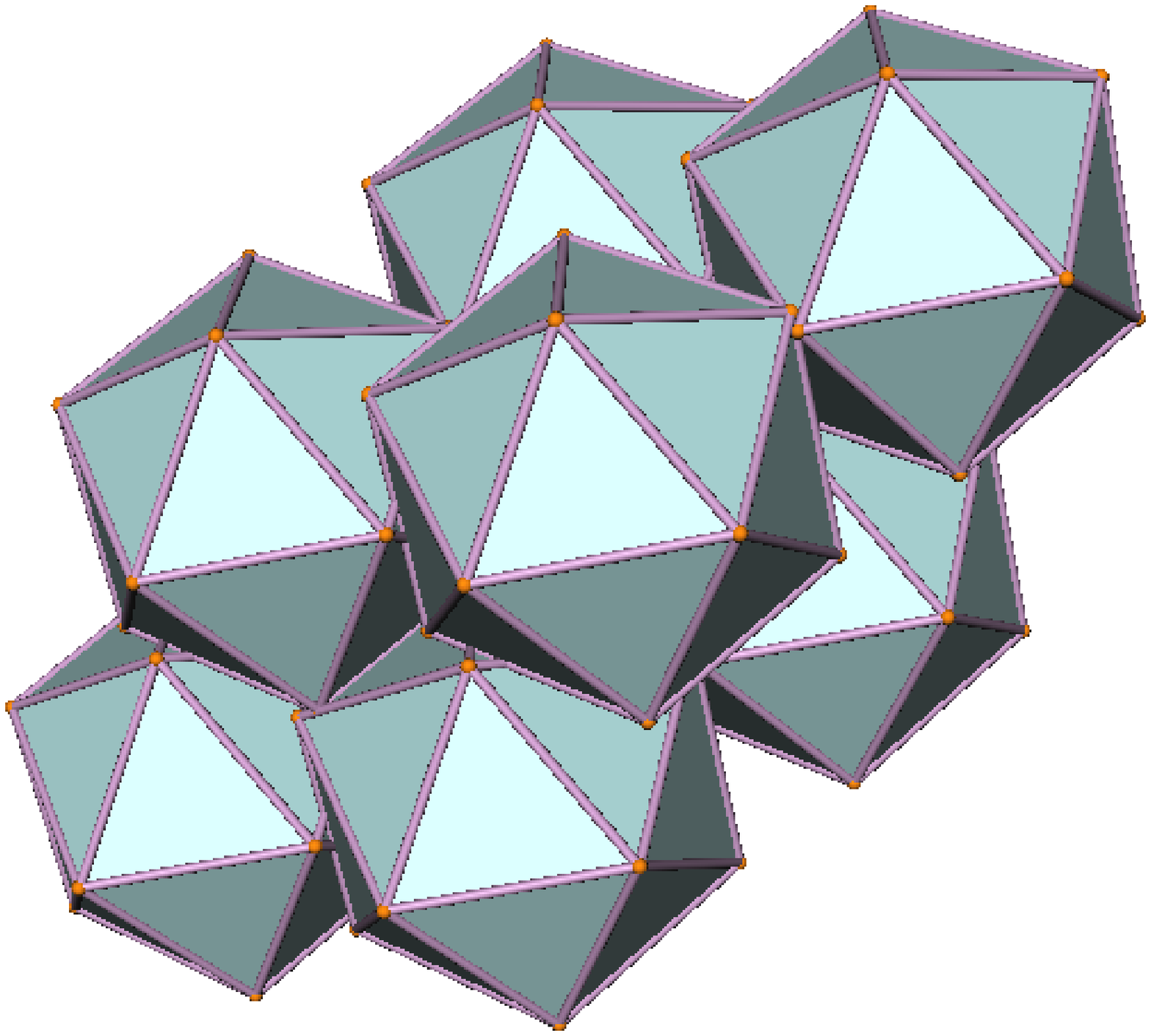} &
\includegraphics[width=4.0cm,keepaspectratio]{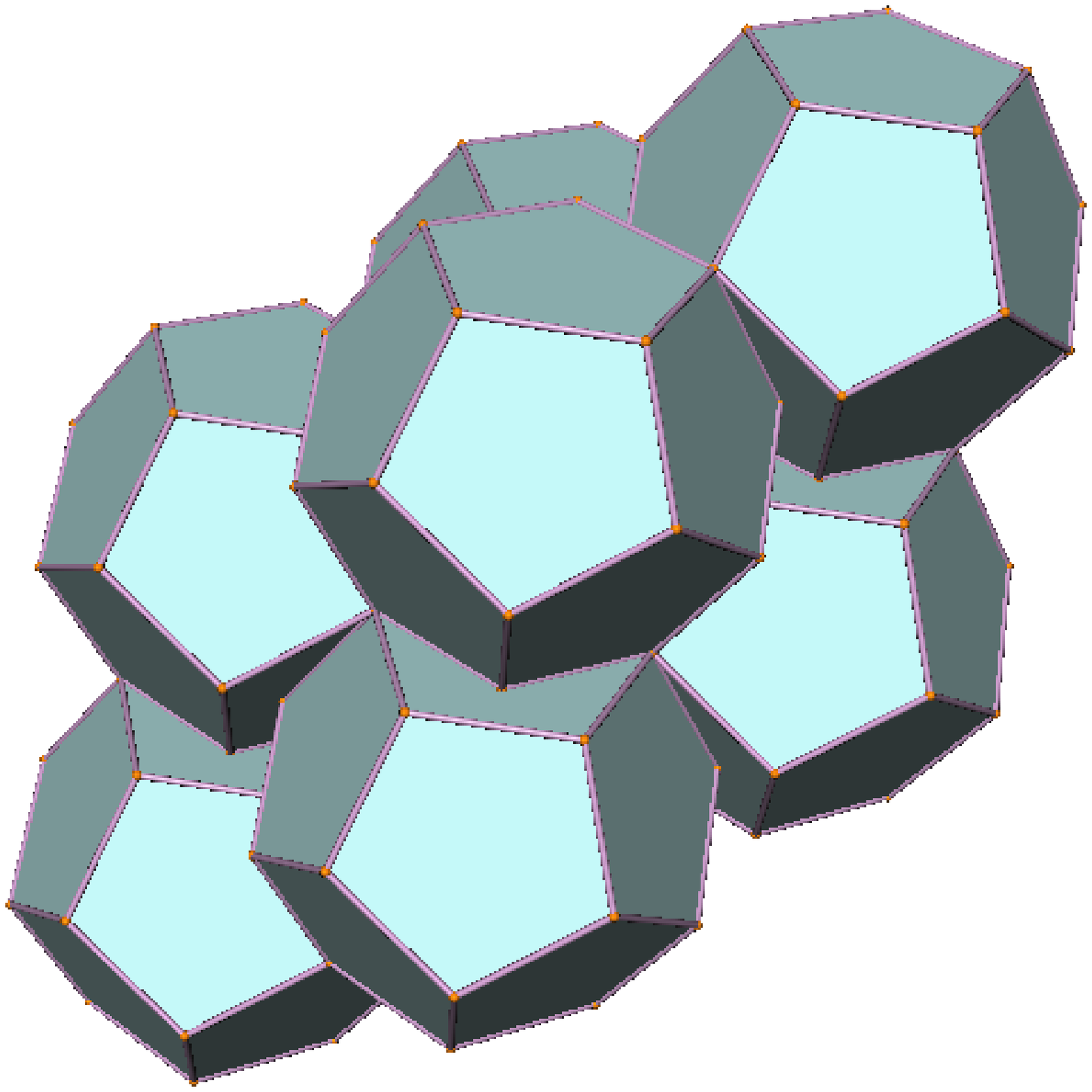} &
\includegraphics[width=4.25cm,keepaspectratio]{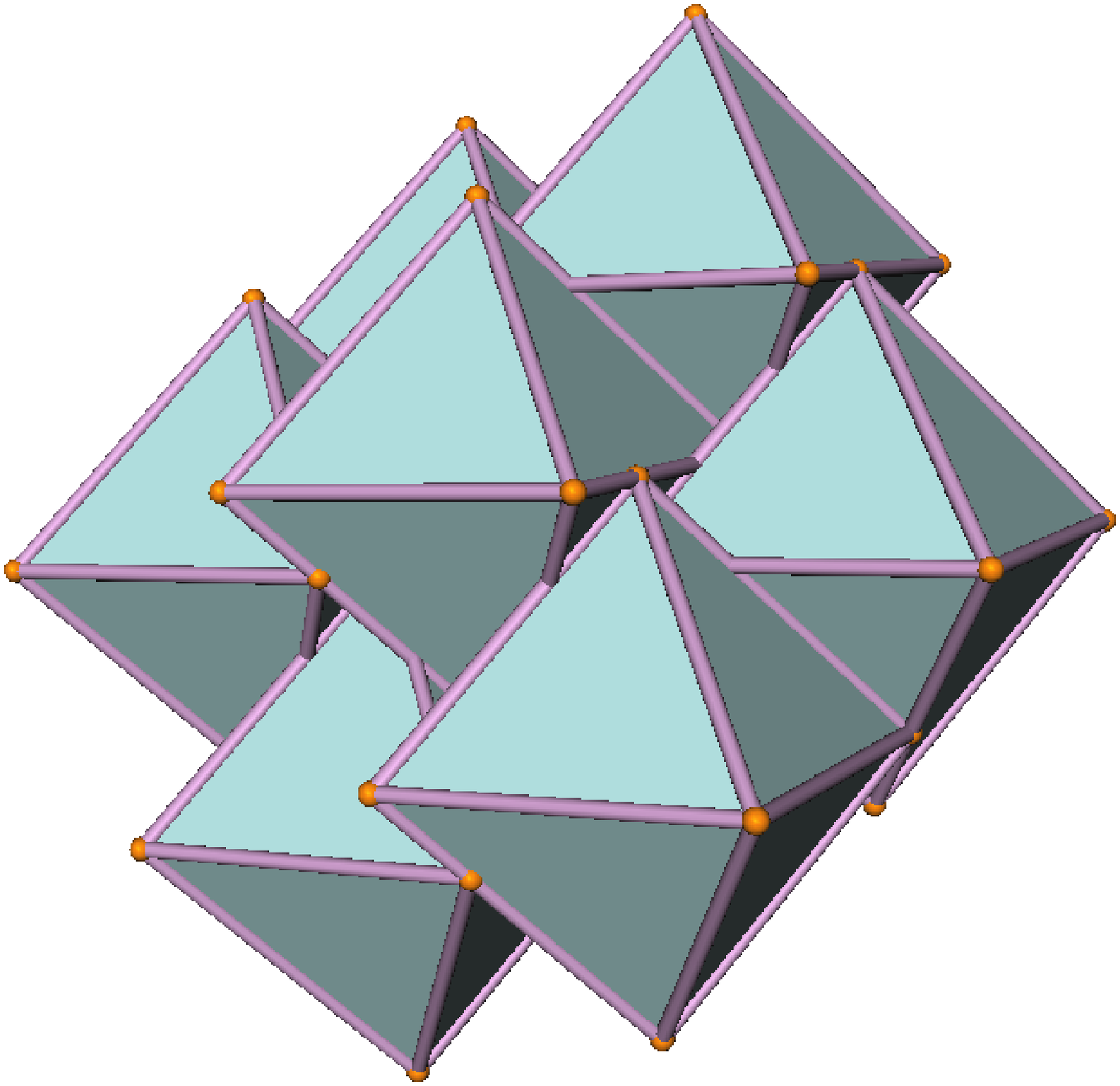} \\
\mbox{\bf (a)} & \mbox{\bf (b)} & \mbox{\bf (c)} & \mbox{\bf (d)}\\
\end{array}$
\end{center}
\caption{Portions of the  densest packing of tetrahedra
that we obtain from our simulations and 
the optimal lattice packings of the icosahedra, dodecahedra,
and octahedra that our simulations converge to. All of these packings
are at least {\it locally} jammed, i.e., each particle cannot be translated or
rotated while fixing the positions and orientations of all the other particles \cite{To01,Do07}.
We emphasize
that even though the latter three cases begin with complex
multi-particle initial configurations in the large fundamental (repeating)
cell, they all converge to packings in which a smaller repeat unit
contains only one centroid, i.e., they all converge to lattice
(i.e., Bravais lattice) packings and, in fact, the corresponding densest lattice
packings.
(a) Tetrahedral packing. We depict the 72 particles in the fundamental
cell of this non-lattice packing. Within the cell, the particles
are characterized by short-range translational order and a preference for 
face-to-face contacts (see Supplementary information).
(b) Optimal lattice packing of icosahedra. (c) Optimal lattice packing of dodecahedra. 
(d) Optimal lattice packing of octahedra.}
\label{tetra}
\end{figure}

To obtain dense packings of  icosahedra, dodecahedra and octahedra,
 we use a wide range of initial configurations.
These include multi-particle configurations (with $N$ ranging from 20 to 343) of random ``dilute" packings and a variety
of lattice packings with a wide range of densities.
In the case of icosahedra, dodecahedra and octahedra, we obtain final packings
with densities at least as large
as $0.836315 \ldots$, $0.904002\ldots$ and $0.947003\ldots$, respectively,
which are extremely close in structure and density to their
corresponding optimal lattice packings with $\phi^{lattice}_{max} = 0.836357\ldots$ \cite{Be00},
$\phi^{lattice}_{max} = (5+\sqrt{5})/8 = 0.904508\ldots$  \cite{Be00},
and $\phi^{lattice}_{max} = 18/19 = 0.947368\ldots$ \cite{Mi04}, respectively. 
Figure \ref{tetra} shows the optimal lattice packings of icosahedra, dodecahedra
and octahedra in which each particle contacts 12, 12  and 14
others, respectively. Our simulation results strongly suggest that the optimal lattice
packings of the centrally symmetric Platonic solids are indeed the densest
packings of these particles, especially since these arise from a
variety of initial ``dilute'' multi-particle configurations within
an {\it adaptive} fundamental cell.

 We can show that the maximal density $\phi_{max}$ of a packing of congruent nonspherical particles of
volume $v_p$ is bounded from above according to
\begin{equation}
\phi_{max}\le \phi_{max}^{upper~bound} = \mbox{min}\left[\frac{v_{particle}}{v_{sphere}}\;
\frac{\pi}{\sqrt{18}},1\right], \label{bound}
\end{equation}
where $v_{sphere}$ is the volume of the largest sphere that can be inscribed
in the nonspherical particle and $\pi/\sqrt{18}$ is the maximal sphere-packing density.
The proof is given in the Supplementary Information.
The upper bound (\ref{bound}) will be relatively tight for packings of nonspherical
particles provided that the {\it asphericity} $\gamma$
(equal to the ratio of the circumradius to the inradius)  of the
particle is not large.  Since bound (\ref{bound}) cannot generally be sharp (i.e., 
exact) for a nontiling, nonspherical
particle, any packing  whose density is close to the upper bound
(\ref{bound}) is nearly optimal, if not optimal.

\begin{figure}
\begin{center}
\includegraphics[height=8.5cm,keepaspectratio]{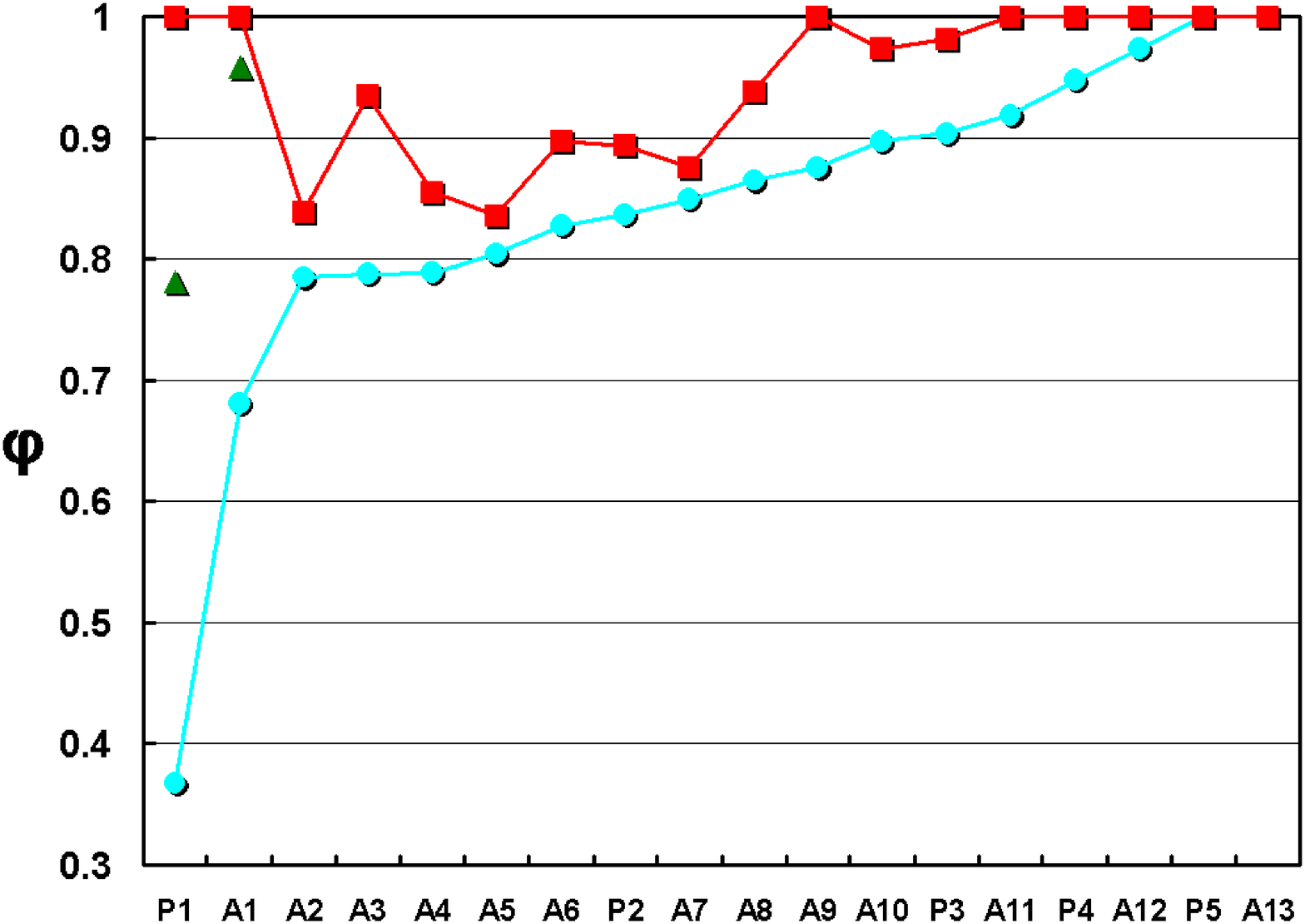}
\end{center}
\caption{Comparison of the densest known lattice packings (blue circles)  of the
Platonic and Archimedean solids \cite{Mi04,Ho70,Be00} to the corresponding upper bounds (red squares)
obtained from (\ref{bound}). The large asphericity
and lack of central symmetry of the tetrahedron (P1) and truncated tetrahedron (A1)
are consistent with the large gaps between their upper-bound densities 
and densest-lattice-packing densities, and the fact that there are non-lattice packings
with densities appreciably greater than $\phi_{max}^{lattice}$ (depicted as green
triangles in the figure).
The truncated tetrahedron is the only non-centrally symmetric Archimedean
solid, the densest known packing, which is a non-lattice
packing with two particles per fundamental cell 
and a density at least as high as $23/24 = 0.958333\ldots$ \cite{Co06}.}
\label{density-bounds}
\end{figure}

Figure \ref{density-bounds} compares the density of the densest lattice
packings of the Platonic and Archimedean solids to the corresponding upper bounds
on the maximal density for such packings. 
The central symmetry of the majority of the Platonic and Archimedean solids
and their associated relatively small asphericities
explain the corresponding small differences between $\phi_{max}^{lattice}$
and $\phi_{max}^{upper~bound}$ and is consistent with our simulation findings
that strongly indicate that their optimal arrangements are their respective
densest lattice packings.

Why should the densest packings of the centrally symmetric solids
be their corresponding optimal lattice packings? 
First, note that face-to-face contacts allow
such polyhedral packings to achieve higher densities
because they enable the contacting centroids around
each particle to come closer together. Second, face-to-face
contacts are maximized when each particle has the same
orientation because of the central symmetry and 
the equivalence of the three principle axes 
(associated with the small asphericity) of the solid.
This is consistent with a lattice packing, the 
densest of which is the optimal one.
These arguments in conjunction with our simulation results and
rigorous bounds lead us to the following conjecture: {\it The densest packings of the centrally symmetric
Platonic and Archimedean solids are given by their corresponding optimal lattice packings.}
This is the analog of Kepler's sphere conjecture for 
these solids. 

There is no reason to believe that
denser packings of tetrahedra cannot be achieved by employing even
better initial conditions than those based on the  wagon-wheels
packing and a larger number of particles.
 Observe that the densest packings of all of the
Platonic and Archimedean solids reported here as well as the
densest known packings of superballs \cite{Ji09} and ellipsoids
\cite{Do04} have densities that exceed the optimal sphere packing
density $\phi^{sphere}_{max}=\pi/\sqrt{18}=0.7408\ldots$. These results
are consistent with a conjecture of Ulam \cite{Ulam}. 
Ulam's conjecture may be violated if the convex particle has little or
no symmetry, but a counterexample has yet to be given.

How does our conjecture extend to other polyhedral packings?
It's natural to group the infinite families of prisms and antiprisms \cite{Cr97} with the
Archimedean solids. A {\it prism} is a polyhedron having bases that are parallel, congruent polygons 
and sides that are parallelograms. An {\it antiprism} is a polyhedron 
having bases that are parallel, congruent polygons  and sides that are alternating bands of triangles.  Prisms with an even number of sides and antiprisms are centrally
symmetric and so it may be that Bravais lattices of such solids are optimal.
However,  prisms with an odd number of sides are not centrally symmetric and 
thus their optimal packings may not be Bravais lattices. In future work,
we will determine whether our conjecture extends to prisms
and antiprisms.

\noindent{\bf \large METHODS SUMMARY}

The objective function in our ASC optimization scheme is
taken to be the negative of the packing density $\phi$. Starting
from an initial packing configuration in the
fundamental cell, the positions and orientations of the polyhedra
are design variables for the optimization. Importantly, we also
allow the boundary of the fundamental cell to deform
as well as shrink or expand such that there is a
net shrinkage (increase of the density) in the
final state. Thus, the deformation and
compression/expansion  of the cell boundary are also design variables. We are
not aware of any packing algorithm that employs both a {\it
sequential} search of the configurational space of the particles
and the space of lattices via an {\it adaptive fundamental cell} that shrinks on average to obtain
dense packings. The ASC has a number of novel features
that distinguish it from previous packing algorithms
that have been devised for spheres \cite{Jo85,Ri98,Uc04}, ellipsoids \cite{Do05a,Do05b}
and superballs \cite{Ji09} (see Methods for details).


\begin{addendum}
 \item We are grateful to Henry Cohn and John Conway
for helpful comments on our manuscript. S. T. thanks the Institute
for Advanced Study for its hospitality during his stay there. This
work was supported by the National Science Foundation under Award Numbers 
DMS-0804431 and DMR-0820341. The figures showing the polyhedra were generated 
using the AntiPrism package developed by Adrian Rossiter.
 \item[Competing Interests] The authors declare that they have no
competing financial interests.
 \item[Correspondence] Correspondence and requests for materials
should be addressed to S.T~(email:\linebreak
torquato@electron.princeton.edu). 
\end{addendum}

\noindent{\bf METHODS}

The ASC optimization problem could be solved using various
techniques, depending on the shapes of the particles. For example,
for spheres,  linear programming (LP) techniques can
efficiently produce optimal solutions (Torquato, S. \& Jiao, Y.).
However, for polyhedra, the complex nonoverlap conditions 
make the ASC scheme inefficient to solve using LP methods.  For polyhedral particles, 
we solve the ASC optimization problem using a standard Monte Carlo (MC) procedure
 with a Metropolis acceptance rule for trial moves to search
the design variable space (DVS) efficiently, 
which contains both the configuration space of the particles and 
the space of lattices.

In our implementation, a polyhedral particle is represented by the 
position of its centroids as well as the coordinates of all its 
vertices \textit{relative} to the centroid. Note that although this 
representation contains redundant information, it is a convenient way to 
deal with the rotational motions of the polyhedra. To search 
the configuration space of the particles, small random trial moves 
of arbitrarily selected particles are attempted sequentially for each particle. Each 
trial move is equally likely to be a translation of the centroid of the particle
or a rotation of the particle about a randomly 
oriented axis through its centroid. 

The space of lattices is searched by deforming/compressing/expanding 
the fundamental cell, which is completely characterized by a 
strain tensor in the linear regime (i.e., small strain limit). 
The trace of the strain tensor determines the volume change of the 
fundamental cell and is involved in the objective function. The 
off-diagonal components of the tensor determines the shape change of the cell. 
The positions of the particles centroids are \textit{relative coordinates} 
with respect to the lattice vectors. When the strain tensor is 
applied to the lattice vectors, although the relative coordinates of the centroids remain 
the same, the \textit{Euclidean distances} between the particles will change. 
Thus, the deformation/compression/expansion of the fundamental cell at least in part
allows for \textit{collective} particle motions, which is more efficient in 
finding a direction in the DVS leading to a higher packing density. 
Moreover, it is the overall compression of
the fundamental cell that causes the packing density to increase,
not the growth of the particles as in most molecular dynamics and MC hard-particle
packing algorithms \cite{Jo85,Ri98,Uc04,Do05a,Do05b,Ji09}. 
It should be noted that for \textit{polyhedral particles}, 
an algorithm that employs particle growth 
with an adaptive {\it non-shrinking} fundamental cell is computationally less efficient 
than the ASC scheme that fixes the particle size while allowing the cell 
to shrink on average.

In the simulation, starting from an initial configuration of polyhedral particles, 
a trial configuration can be generated by moving (translating
and rotating) a randomly chosen particle or by a random 
deformation and compression/expansion of the fundamental cell. 
If this causes interparticle overlaps, the trial configuration is rejected; 
otherwise, if the fundamental cell shrinks in size (which makes the density $\phi$ higher), 
the trial configuration is accepted.
On the other hand,  if the cell expands in size, 
the trial configuration is accepted with a specified probability $p_{acc}$, which 
is made to decrease as $\phi$ increases and approaches zero at the jamming 
limit \cite{To01} (i.e., locally maximally dense packing) is reached. In particular, we find 
$p_{acc}$, with an initial value $p_{acc}\sim 0.35$ and decreasing as a power law, works well for 
most systems that  we studied. 
The ratio of the number of particle motions to the number of cell trial moves 
should be greater than unity (especially towards the end of the simulation), 
since compressing a dense packing could
cause many overlaps between the particles. Depending on the initial configuration, 
the magnitudes of the particle motions and the strain components need to be 
chosen carefully to avoid the system getting stuck in some shallow local minimum.

A crucial aspect of any packing algorithm is the need to check for interparticle
overlaps under attempted particle motions. Hard polyhedron particles, 
unlike spheres, ellipsoids and superballs, 
do not possess simple ``overlapping" functions. (The overlap function of a pair of strictly convex and
smooth particles is a function of the positions, orientations and
shapes of the two particles, whose value indicates whether the two particles
overlap or not, or whether they are tangent to one another.) The {\it separation axis 
theorem} \cite{Ra94} enables us to check for interparticle overlaps for polyhedra up to the 
numerical precision of the machine.
In particular, the theorem states that two convex polyhedra 
are separated in space if and only if there exists
an axis, on which the projections of the vertices of 
the two polyhedra do not overlap. The separation axis is either
perpendicular to one of the faces of the polyhedra 
or perpendicular to a pair of edges from different polyhedra. 
Thus, this reduces the  number of axes that need to be checked  
from infinity to $[E(E-1)/2+2F]$, 
where $E$ and $F$ is the number of edges and faces of the polyhedra,
respectively. A pre-check using the circumradius and inradius 
of the polyhedra could dramatically speeds up the
simulations, i.e., two particles are guaranteed to overlap if the centroidal
separation is smaller than twice the inradius and guaranteed
not to overlap if the centroidal separation is larger than 
twice the circumradius. The {\it circumsphere} is the smallest sphere containing the particle.
The {\it insphere} is the largest sphere than can be inscribed
in the particle. 

The cell method and near-neighbor list \cite{Do05a, Do05b} 
are also employed to improve the efficiency of the simulation, 
but are appropriately modified to incorporate the adaptive 
fundamental cell.


\begin{thebibliography}{30}

\bibitem{Ber65}
Bernal, J. D. in {\it Liquids: Structure, Properties, Solid
Interactions} (eds Hughel, T. J.)
 25-50 (Elsevier, 1965).

\bibitem{Za83}
Zallen, R. {\it The Physics of Amorphous Solids} (Wiley, 1983).

\bibitem{To02}
Torquato, S. {\it Random Heterogeneous Materials: Microstructure
and Macroscopic Properties} (Springer-Verlag, 2002).

\bibitem{Ch00}
Chaikin, P. M.\& Lubensky, T. C. {\it Principles of Condensed
Matter Physics} (Cambridge University Press, 2000).


\bibitem{Ed94}
Edwards, S. F. in {\it Granular Matter} (eds Mehta, A.) 121-140
(Springer-Verlag, 1994).

\bibitem{Li01}
Liang, J. \& Dill, K. A. Are proteins well-packed? {\it Biophys J.} {\bf 81}, 751-7666 (2001).

\bibitem{Pu03}
Purohit, P. K., Kondev, J. \& Phillips, R. Mechanics of DNA packaging in viruses.
{\it Proc. Nat. Acad. Sci.} {\bf 100}, 3173-3178 (2003).

\bibitem{Ge08}
Gevertz, J. L. \& Torquato, S. A Novel Three-Phase Model of Brain Tissue Microstructure. PLoS Comput. Biol. {\bf 4}, e1000152 (2008).

\bibitem{Co98}
Conway, J. H. \& Sloane, N. J. A. {\it Sphere Packings, Lattices
and Groups} (Springer-Verlag, 1998).

\bibitem{Ha05}
Hales, T. C. A proof of the Kepler conjecture. {\it Ann. Math.}
{\bf 162}, 1065-1185 (2005).

\bibitem{Co03}
Cohn, H. \& Elkies, N. New upper bounds on sphere packings I. {\it
Ann. Math.} {\bf 157}, 689-714 (2003).

\bibitem{Do04}
Donev, A., Stillinger, F. H., Chaikin, P. M. \& Torquato, S.
Unusually dense crystal ellipsoid packings. {\it Phys. Rev. Lett.}
{\bf 92}, 255506 1-4 (2004).

\bibitem{Co06}
Conway, J. H. \& Torquato, S. Packing, tiling and covering with
tetrahedra. {\it Proc. Nat. Acad. Sci.} {\bf 103}, 10612-10617
(2006).

\bibitem{Ch08}
Chen, E. R. A dense packing of regular tetrahedra. {\it Discrete
Comput. Geom.} {\bf 40}, 214-240 (2008).

\bibitem{Ji09}
Jiao, Y., Stillinger, F. H. \& Torquato, S. Optimal packings of
superballs. {\it Phys. Rev. E} {\bf 79}, 041309 1-12 (2009).

\bibitem{Ra94}
Golshtein, E. G. \& Tretyakov, N. V. \textit{Modified Lagrangians
and Monotone Maps in Optimization} (Wiley, 1996).

\bibitem{Ho70}
Hoylman, D. J. The densest lattice packing of tetrahedra. {\it
Bull. Am. Math. Soc.} {\bf 76}, 135-137 (1970).

\bibitem{To01}
Torquato, S. \& Stillinger, F. H. Multiplicity of generation,
selection, and classification procedures for jammed hard-particle
packings. {\it J. Phys. Chem. B} {\bf 105}, 11849-11853 (2001).

\bibitem{Do07}
Donev, A., Connelly, R., Stillinger, F. H. \& Torquato, S.
Underconstrained jammed packings of nonspherical hard particles:
ellipses and ellipsoids. {\it Phys. Rev. E} {\bf 75}, 051304 1-32
(2007).

\bibitem{Be00}
Betke, U. \& Henk, M. Densest lattice packings of 3-polytopes.
{\it Comput. Geom.} {\bf 16}, 157-186 (2000).


\bibitem{Mi04}
Minkowski, H. Dichteste gitterf{\"o}rmige Lagerung kongruenter
K{\"o}rper. {\it Nachr. K. Ges. Wiss. G{\"o}ttingen,
Math.-Phys. KL} 311-355 (1904).



\bibitem{Ulam}
Gardner, M. {\it The Colossal Book of Mathematics: Classic
Puzzles, Paradoxes, and Problems} (Norton, 2001).

\bibitem{Cr97}
Cromwell, P. R. {\it Polyhedra} (Cambridge University Press, 1997).

\bibitem{Jo85}
Jodrey, W. S. \& Tory, E. M. Computer simulation of close random
packing of equal spheres. {\it Phys. Lett. A} {\bf 32}, 2347-2351
(1985).

\bibitem{Ri98}
Rintoul, M. D. \& S. Torquato, S. Hard-sphere statistics along the
metastable amorphous branch. {\it Phys. Rev. E} {\bf 58}, 532-537
(1998).

\bibitem{Uc04}
Uche, O. U., Stillinger, F. H. \& Torquato, S. Concerning maximal
packing arrangements of binary disk mixtures. {\it Physica A} {\bf
342}, 428-446 (2004).



\bibitem{Do05a}
Donev, A., Torquato, S. \& Stillinger, F. H. Neighbor list
collision-driven molecular dynamics for nonspherical hard
particles: I. algorithmic details. {\it J. Comput. Phys.} {\bf
202}, 737-764 (2005).

\bibitem{Do05b}
Donev, A., Torquato, S. \& Stillinger, F. H. Neighbor list
collision-driven molecular dynamics for nonspherical hard
particles: II. applications to ellipses and ellipsoids. {\it J.
Comput. Phys.} {\bf 202}, 765-793 (2005).



\end{thebibliography}
\end{document}